\documentstyle[aps,epsf,multicol,psfig]{revtex}
\def\iv{{\it in vitro }}
\def\x{\vec{x}}

\def\bfg{\begin{figure}}
\def\efg{\end{figure}}
\def\be{\begin{equation}}
\def\ee{\end{equation}}

\begin{document}
\pssilent
\title{Exponential Distribution of Locomotion Activity in Cell
Cultures}

\author{Andr\'as Czir\'ok$^1$, Katalin Schlett $^2$, Em\'\i lia
Madar\'asz $^2$,
and Tam\'as Vicsek$^1$}

\address{
$^1$  Department of Biological Physics,
E\"otv\"os University, Budapest, Puskin u. 5-7, 1088 Hungary \\
$^2$ Department of Comparative Physiology,
E\"otv\"os University, Budapest, M\'uzeum krt. 4/A, 1088 Hungary \\
}

\date{\today}
\maketitle

\begin{abstract}

{\it In vitro} velocities of several cell types have been measured using
computer controlled video microscopy, which allowed to record the cells'
trajectories over several days. On the basis of our large data sets we show
that the locomotion activity displays a {\it universal} exponential
distribution.  Thus, motion resulting from complex cellular processes can be
well described by an unexpected, but very simple distribution function. A
simple phenomenological model based on the interaction of various cellular
processes and finite ATP production rate is proposed to explain these
experimental results.
\end{abstract} 

\begin{multicols}{2}

Tissue cell migration \cite{migration} has been studied in a number of 
biological processes, including the development of the nervous system
\cite{neu}, immune reactions \cite{NKKKC96} and tumor
spreading \cite{CS95}.  Beside the inherent biological importance, cell
locomotion plays a key role in a number of interesting large scale pattern
formation phenomena like cell sorting \cite{Foty94,MG96,RKS98}, or aggregation
of amoebae \cite{dict}. Although the
behavior of swimming bacteria has been well studied \cite{berg} and served as a
basis to understand collective microbiological phenomena \cite{coll_baci},
rather little is known about the basic statistical features of eukaryotic cell
migration on time scales comparable to the time of the pattern formation
processes.

The locomotion of cells emerges from parallel cyclic subprocesses
including cell-attachment to and detachment off the substrate, growth and
collapse of filopodia/lamellopodia, the dynamic maintenance of appropriate cell
surface composition and displacement of the cell mass.  This complexity
suggests two consequences.  On one hand, strong correlation cannot be expected
between the actual cell velocity and the concentration of a single (or a few)
cell biological regulatory factor(s) \cite{MGM95}.  On the
other hand, interacting complex units often results a characteristic
macroscopic behavior which can be understood by mathematical models
\cite{Foty94,MG96,RKS98,dict,coll_baci}, thus some sort of statistical
regularities might be expected in the cellular behavior as well.

In order to obtain data providing a basis for an extensive statistical analysis
we have developed a computer controlled system which automatically collects
data and maintains the cell culture conditions in a chamber attached to the
microscope for several days.  We found that the locomotion activity displays an
{\it exponential distribution}, i.e.,   in spite of the complex cellular
processes, the motion of a cell can be well described by a simple universal
distribution function.   This observation can be explained in terms of a
phenomenological model based on the competition of various cellular processes
for the finite free energy (ATP) resources.

With the aim of applying statistical approaches in the study of cellular
locomotory activity, long-term cell migration patterns were recorded in
low-density monolayer cultures (see Table 1).   The cellular trajectories (Fig.
1) were tracked using a computer controlled phase contrast videomicroscope
system.  Due to a high degree of automatization we could collect and analyze a
large set of data (about 100,000 cell positions), significantly exceeding the
amount of information evaluated in previous works on cell motility 
\cite{neu,MG96,RKS98,DB87,PS92,FMH93,BPI94,HTMR96}.

Average velocities were calculated over a time period of one hour ($t_0$) as
$v_i(t)=\vert \x_i(t+t_0/2)-\x_i(t-t_0/2)\vert/ t_0$, where the location of the
$i$th cell  at time $t$ is denoted by $\x_i(t)$.  Remarkable fluctuations were
found in the locomotion activity of individual cells, a phenomenon
which has also been reported in living slices of developing cortex
\cite{neu}, in vitro migration of neurons\cite{neuron} and fibroblasts
\cite{DB87}.  The characteristic time scale of the fluctuations was
estimated to be a few hours, a value similar to that reported for
fibroblasts \cite{DB87}.

Due to the low cell density, most cells migrated freely, but the effects
of various cell interactions could be also investigated.  Since it is known
that high cell density can significantly decrease the cell locomotion activity
\cite{contact}, we investigated whether the observed velocity fluctuations were
caused by changes in local intercellular connections.  Cell to cell contacts
were identified on the images and the locomotion velocities $v_i(t)$ were
sorted into two sets (``solitary'' / ``interacting'') depending on whether the
cell in question had or had not visually observable contacts with
other cells.  Cumulative velocity distribution functions $F(v)$ were calculated
for time periods of $20$ hours giving the probability of the relation
$v_i(t)<v$ to hold for a randomly selected cell $i$ and a random time $t$.
Comparing $F(v)$ of ``solitary'' cells to that of ``interacting'' cells (Fig.
2a), two almost overlapping ranges of velocities ($0-80$ $\mu$m/h and $0-100$
$\mu$m/h, respectively) were found.   It is also known \cite{migration} that
during migration cells leave various extracellular molecules and other
polipeptides attached to the substrates.  To elucidate the effect of these cell
trails on the velocity fluctuations we filtered our database in such a way that
we kept only those cell positions which distance was greater than $20 \mu$m
(i.e., twice the typical cell size) of the trajectory of other cells.   Again,
no qualitative difference was found in the velocity fluctuations suggesting
that they are {\it inherent property of the cells}.

One of our main findings is that $F(v)$ of all cultures investigated
could be very well fitted by an exponential distribution
\be
F(v)=1-e^{-\alpha v}
\label{exp}
\ee
within measurement errors in the {\it entire} velocity range studied (Fig. 2b).
This result indicates that the exponential velocity distribution is likely to
be a general characteristic feature of \iv cell motility for a broad class  of
surface attached cells. 

The above velocity distributions were found for relatively large ($\approx 100$
cells) populations.  On the level of individual cells the exponential behaviour
(\ref{exp}) can be interpreted in two ways: (i) The culture is
inhomogeneous, i.e., slower and faster cells can be distinguished on the bases
of well preserved phenotypic properties.  In this case the exponential $F(v)$
distribution can reflect the ratio of the slow and fast cells in the culture,
while the velocity fluctuations of the individual cells can show an arbitrary
distribution.  (ii) If the culture is homogeneous, then almost all cells exhibit
the same distribution of velocity fluctuations, i.e., $F_i(v)\approx F(v)$
holds for each cell $i$, where $F_i(v)$ denotes the distribution function of
the cell $i$. In this scenario the average velocity $\langle v_i(t) \rangle$ of
each cell would be the same if we could calculate the time averages over an
infinitely long time. Since the time averages are calculated over a finite time
$\tau$ only, for the distribution of the average velocities we can expect a
Gamma distribution with a parameter $s=\tau/t_0$, where the correlation time of
the process is denoted by $t_0$ \cite{noteI}.

To decide between the above alternatives $F_i(v)$ and the average velocity
$\langle v_i(t) \rangle$ over a period of $\tau=16$h were calculated for each
cell $i$. Fig. 3a and 3b show that the experimental data clearly support the
homogeneous alternative.  As an example, the data shown in Fig. 3b
can be well fitted by a normalized Gamma distribution $F_\Gamma(u,s)=\int_0^udz
s^s z^{s-1}e^{-sz}/(s-1)!$ with a parameter value of $s=9$.   The fitted value
of $s$ is consistent with the finding that the correlation time scale of the
velocity fluctuations is in the order of hours.

The fact that the distribution of cellular velocities follows a {\it simple}
exponential function is an unexpected finding.  Several subprocesses of cell
locomotion (receptor binding, membrane exocitosis, lamelopodium/filopodium
formation, etc.) display either Poisson distribution \cite{BPI94} or periodic
behavior \cite{HB97} if studied on time scales  short enough to exclude
non-stationarities of the cell state. However, the fluctuations of cell
velocities recorded in our experiments were observed on significantly longer
time scales than those listed above. Since in a few hours many lamellopodia are
formed and many vesicles are fused to the membrane, etc., the relative
frequency fluctuation of such stationary processes must be small and close to 
Gaussian.  In agreement with this expectation, previous works on cell
locomotion -- although not calculating $F(v)$ explicitly -- have set up
phenomenological models predicting Gaussian velocity distribution
\cite{DB87,PS92,BPI94}.  Also, a close to Gaussian distribution was found in
\cite{MG96} where motion within a cell aggregate was studied. In that case,
however, there was a strong interaction between the cells and the displacement
of a single cell was also a result of the activity of the neighboring cells.  

Within the time scale of the velocity fluctuations both the cellular
environment and the pattern of gene expression in the cells can be considered
as stationary.   This time, on the other hand, is long enough to permit changes
in cell motility as a consequence of changes in the concentration of several
intracellular regulatory factors.  In the following we demonstrate with a very
simple cell model that the exponential distribution of cell velocity
fluctuations can indeed reflect such stochastic intracellular changes.

Since the cell locomotion processes are cyclic and dissipative, the observed
velocity is proportional to the amount $J_0$ of free energy dissipated (number
of ATP molecules hydrolized) in the appropriate chemical reactions contributing
to cell motility:
\be
v(t)=\mu J_0(t),
\label{E0}
\ee
where $\mu$ denotes the conversion factor between velocity and energy
dissipation. It is natural to assume that $\mu$ is constant, i.e., the
subprocesses are synchronized in such a manner that a close to optimal
distribution of the energy inflow is maintained.

The individual reaction rates are determined on one hand by the ATP
concentration $c$ and on the other hand by the concentrations of various
regulatory factors.  Some of these factors maintain the motility efficiency
$\mu$, while other factors are products of different cellular processes, with
the potential of altering the cell locomotion activity.  These latter factors
can change {\it independently} from each other: while some of the changes
increase, some others decrease the cell motility. Thus, we assume that the
temporal changes of $J_0$ can be represented as
\be
{dJ_0\over dt}=g_0{dc\over dt}+\xi_0,
\label{E2a}
\ee
where $g_0(t)>0$ and $\xi_0(t)$ is a stochastic variable with zero mean and
Gaussian distribution.

Finally, we assume that the cellular ATP production rate $P$ is limited 
by the ATP production capacity of the cell, so the various cellular processes
(among them cell locomotion, protein synthesis, DNA duplication, etc.) compete
for the ATP available.  Since the changes of $c$ can be calculated as a
difference in the production and the consumption rates, the equations
\be
{dJ_\ell\over dt}=g_\ell(P-\sum_{k=0}^N J_k)+\xi_\ell ~~~~~~~~\ell=0,1,...,N
\label{E4}
\ee
are obtained, where $J_\ell$ denotes the rate of free energy dissipation in a
given cellular process $\ell$, while $g_\ell$ and $\xi_\ell$ are  analog
quantities to $g_0$ and $\xi_0$, respectively.

\def\J{\vec{J}}

It can be seen that Eqs. (\ref{E4}) describe a diffusion process in the
vicinity of the $\sum_{k=0}^N J_k=P$ hyperplane.  Accordingly, the probability
distribution of the ``cell state'' $\J=(J_0,J_1,...,J_N)$ on this hyperplane is
uniform.  Integrating the distribution of $\J$ and taking into account Eq.
(\ref{E0}) we obtain the experimentally observed exponential behaviour
\be
F(v)=1-\Bigl(1-{v\over \mu P}\Bigr)^N \approx 1 - e^{-v/\mu p},
\label{E6}
\ee
where $p=P/(N+1)$ is the average free energy consumption rate of a given
cellular process.

There is a formal similarity between the distribution given by Eq. (\ref{E6})
and the Boltzmann energy distribution within systems in thermal equilibrium. In
the latter case, the uniform microcanonical distribution can be derived using
either quantum mechanics or Liouville's theorem for Hamiltonian systems while
here it is an interesting feature of the interacting regulatory processes.

Although Eq. (\ref{E4}) is certainly oversimplified and the free energy
dissipation rates $J_\ell$ are not identified with actual biochemical
reactions, the exponential behavior of Eq. (\ref{E6}) is expected to hold for a
wide class of models (e.g., non-constant $P$, considering explicitly the
interactions of regulatory networks, etc.).  Our simple calculation also
predicts that exponentially distributed fluctuations can be expected in
numerous other cellular activities as well.

We thank Z. Csah\'ok and O. Haiman for discussions, Zs. K\"ornyei
and B. Heged\H us for help in the data analysis, I. Fazekas for
providing us the human glioblastoma cell lines. Supporting grants OTKA
T019299, F017246 and FKFP 0203/1997 were substantial for our work.

\def\etal{{\it et al.~}}

\end{multicols}

\bfg
\caption{Cellular trajectories of NE-4C cells in a field of $1110$
$\mu$m$\times$ $840$ $\mu$m, during the first $20$ hours in culture.
Microscopic images were taken with $10\times$ objective magnification every 3-4
minutes.  The cells in each 3rd image were tracked manually with a
precision of ca. $5\mu$m, which is comparable with the size of the cells ($10$
$\mu$m). }
\label{fig1}
\efg


\bfg
\caption{ (a) Cumulative velocity distribution functions $F(v)$ calculated from
the trajectories of Fig. 1. for three subsets of cells: ``interacting'',
``solitary'' and cells which do not cross the trails left on the substrate (see
text for details). The similarity of the curves indicates that the fluctuations
are an inherent property of the cells.  The horizontal bars correspond to the
error of velocity determination due mainly to the method of cell positioning.
The vertical bars mark the systematic error ( $\approx 5\%$ ) of $F(v)$, which
estimation was based on the difference between the distribution functions
calculated for the first and the second $10$ hours of the record. The solid
line is a fitted exponential distribution function.  The inset shows the same
distribution functions on a linear-logarithmic plot.  (b) The velocities
$u=v/\overline{\langle v_i(t)\rangle}$ normalized by the average velocity of
the population show an exponential distribution in all the 13 cultures
investigated, demonstrated by the linear region covering four decades in the
linear-logarithmic plot.  The figure shows a typical result for each cell
line.} \label{fig3} \efg

\bfg
\caption{(a) The average function of the single-cell velocity distributions
$\overline{F_i}(v)=\sum_{i=1}^nF_i(v)/n$ show the same exponential
behaviour as the cumulative distribution function $F(v)$ of the whole
cell population shown in Fig.1.
The dotted lines are an exponential fit of $F(v)$
in the corresponding velocity regime. (b) The distribution function
of the normalized average cellular velocities
$\langle u_i \rangle_\tau=\int_0^\tau v_i(t)dt/(\tau \overline{\langle \
v_i(t)\rangle}$) is presented, both as linear and lin-log plots. The solid line
is a fitted Gamma distribution.
}
\label{fig4}
\efg

\begin{table}
\caption{The most important features of the cultures investigated}
\begin{tabular}{|c|c|c|c|c|c|}
Culture  & reference & cell density\tablenote{at seeding; 
$^b$American Type Culture Collection;
$^c$National Institute of Neurosurgery, Hungary;
$^d$migration from cell aggregates} & duration & positions & $\overline{\langle v_i(t)\rangle}$ \\ 
 & & [cells/mm$^2$] & [h] & & [$\mu$m/h] \\ \tableline\tableline
mouse neuroepithelial progenitor (NE-4C)  & [12] & 20, 40, 200 & 24, 24, 12.5 & 2000, 4000, 12500 & 6.7, 8.7, 17 \\ \tableline
mouse fibroblast (NIH-3T3) & ATCC$^b$  & 170 & 20 & 5000 & 12.1 \\ \tableline
human glioblastoma (HA) & OITI$^c$ & 48, 65 & 50, 50 & 13500, 19500 & 11.1, 11.6 \\ \tableline
human glioblastoma (HC) & OITI & 126, 80  & 50, 50 & 37800, 24000 & 6.5, 7.0 \\ \tableline
human breast adenocarcinoma (MDA MB231) & ATCC, [13] & 40 & 24 & 2500 &5 \\ \tableline
rat primary astroglia & [14]&  200$^d$ & 50, 50, 36, 30 & 3500, 3500, 3000, 2500 & 8 \\ 
\end{tabular}
\end{table}


\begin{references}
\bibitem{noteI} The time averaged velocity can be written as $\langle v_i(t) \rangle_{\tau} =
\sum_{t=0}^\tau v(t)/\tau = (1/s) \sum_{a=1}^s (1/t_0) \sum_{t=at_0}^{(a+1)t_0}
v(t)$, where $s=\tau/t_0$.  Since the terms in
the second sum are positively correlated, we estimate the average by $
\langle v_i(t) \rangle_{\tau} \approx (1/s)\sum_{a=1}^sv(at_0)$, which
expression is a sum of $s$ independent, exponentially distributed random
variable.
\end{references}
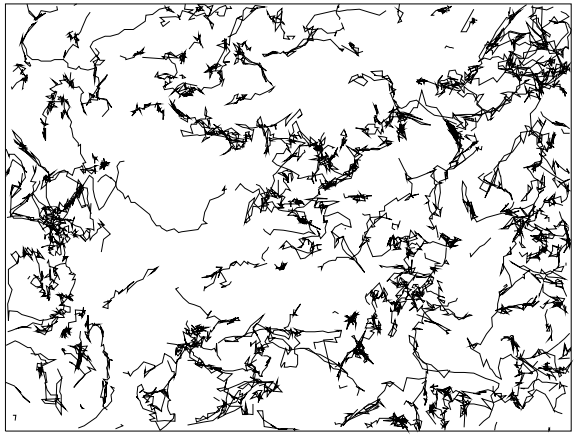
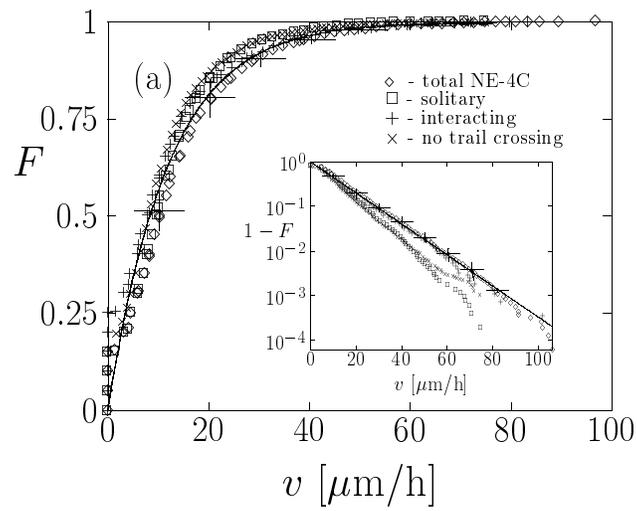
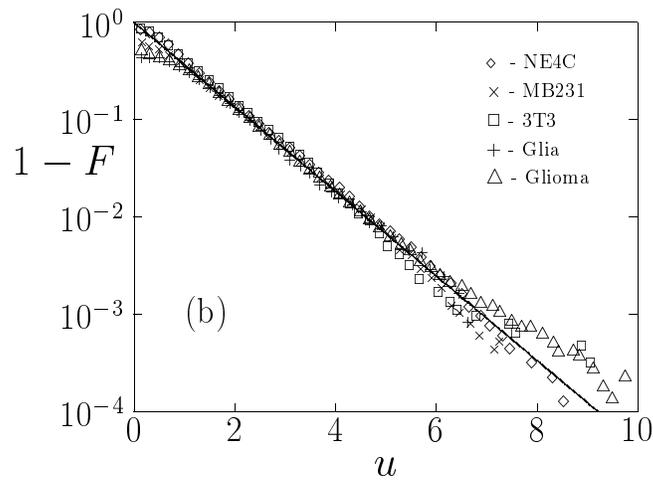
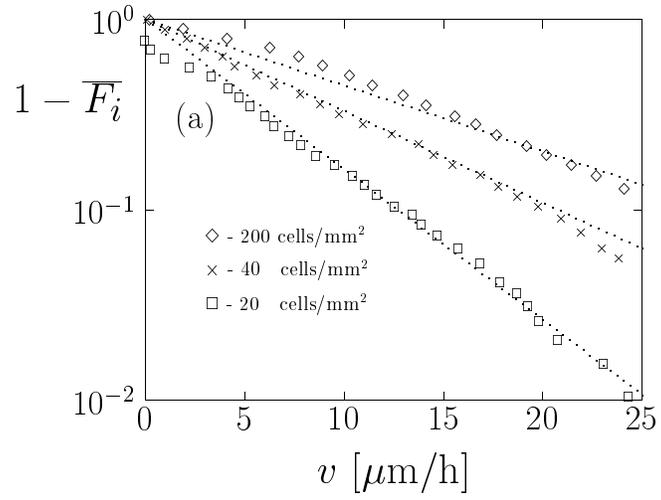
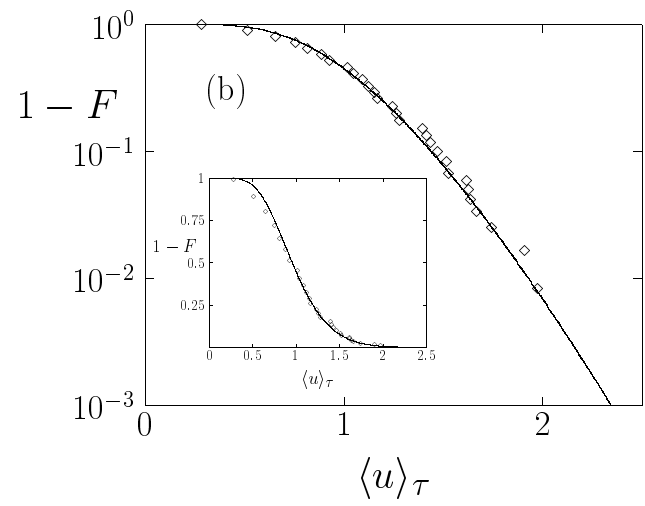
\end{document}